\documentclass[12pt]{article} 
\usepackage{amsmath,amsfonts, amssymb}
\usepackage{comment}
\usepackage{graphicx}
\usepackage{amssymb}
\usepackage{epstopdf}
\newcommand{\be}{\begin{equation}}
\newcommand{\ee}{\end{equation}}
\newcommand{\ba}{\begin{eqnarray}}
\newcommand{\ea}{\end{eqnarray}}
\renewcommand\d{\partial}
\begin{document}

\begin{flushright}
INT-PUB-11-011
\end{flushright}
\begin{flushright}
UCB-PTH-11/01
\end{flushright}
\vskip   1cm
\begin{center} {\Large{\textsc{ Hall viscosity from gauge/gravity duality} }}
\end{center}
\vskip 1cm

\renewcommand{\thefootnote}{\fnsymbol{footnote}}
\centerline{\bf Omid Saremi$^a$\footnote{omid.saremi@berkeley.edu} 
and Dam Thanh Son$^b$\footnote{son@phys.washington.edu}}
\vskip 2cm
\centerline{ ${}^a $\small Berkeley Center for Theoretical Physics and Department of physics, }
\centerline{\small University of California, Berkeley, CA 94720}
\centerline{\small and}
\centerline{\small Theoretical physics group, Lawrence Berkeley National Laboratory, Berkeley, CA 94720}
\bigskip
\centerline{${}^b$\small Institute for Nuclear Theory, University of Washington,}
\centerline{ \small  Seattle, WA 98195 }
\bigskip
\begin{abstract}
In (2+1)-dimensional systems with broken parity, there exists yet
another transport coefficient, appearing at the same order as the
shear viscosity in the hydrodynamic derivative expansion. In condensed
matter physics, it is referred to as ``Hall viscosity''.  We consider
a simple holographic realization of a (2+1)-dimensional isotropic
fluid with broken spatial parity. Using techniques of fluid/gravity
correspondence, we uncover that the holographic fluid possesses a
nonzero Hall viscosity, whose value only depends on the near-horizon
region of the background.  We also write down a Kubo's formula for the
Hall viscosity. We confirm our results by directly computing the Hall
viscosity using the formula.
\end{abstract}
\section{Introduction}
There have been intensive efforts to use methods of gauge/gravity
correspondence~\cite{Maldacena:1997re,Gubser:1998bc,Witten:1998qj} in
studying strongly interacting systems at finite temperature and/or
chemical potential.  One of the motivations for such efforts is to
understand the state of strongly interacting quark-gluon plasma
created at RHIC, which is known to have a shear viscosity to entropy
density ratio not too far from the AdS/CFT value~\cite{Kovtun:2004de}.
The hydrodynamic limit of AdS/CFT has attracted much attention in
recent years. The emerging picture out of these investigations is that
the dynamics of perturbations of a black-brane background is governed,
in the long-wavelength limit, by the same hydrodynamic equations which
describe relativistic fluids~\cite{Shiraz}.

Phenomenologically, the hydrodynamic equations are written down based
on general principles like symmetries and the second law of
thermodynamics as well as the kinetic approach based on the Boltzmann
equation. While the hydrodynamic equations have been known for a long
time, the ability to derive them from holography provides fresh
perspectives, complementary to those of the phenomenological
approach. In particular, holography was instrumental in the discovery
of new hydrodynamic effects in systems with triangle
anomalies~\cite{Shiraz,Erdmenger:2008rm}.  It has been directly
observed that in these systems, there exist additional terms in the
currents of conserved charges, proportional to the vorticity of the
fluid flow. It was further discovered that such contributions are
required by the triangle anomalies and the second law of
thermodynamics, and hence are not restricted to theories with
a gravitational dual~\cite{Son:2009tf}. There have been some attempts to
rederive these terms in the kinetic approach~\cite{Pu:2010as}.

In this paper, we investigate parity-odd effects in (2+1)-dimensional
relativistic hydrodynamics. In contrast to (3+1)-dimensions, there are no
triangle anomalies. However parity may be broken explicitly or
spontaneously. One can ask whether there are new
hydrodynamic effects which are disallowed in parity-invariant fluids. 

It is easy to observe that indeed such an effect does exist, and is
the relativistic generalization of what is called the ``Hall
viscosity'' in the condensed matter literature~\cite{Avron,Avron-JSP}.
In condensed matter, there are variety of models where the Hall
viscosity phenomenon is investigated~\cite{ReadRezayi}. One may ask
whether Hall viscosity appears more generally in a relativistic
context. Consider the stress tensor of a relativistic fluid 
\begin{equation}
  \tau^{\mu\nu} = (E+p)u^{\mu}u^{\nu}+p g^{\mu\nu}
  -\eta P^{\mu\alpha} P^{\nu\beta}V_{\alpha\beta}
  ,\qquad V_{\alpha\beta} \equiv
  \nabla_\alpha u_\beta + \nabla_\beta u_\alpha 
  - g_{\alpha\beta} \nabla\cdot u,
\end{equation}
where $u^\mu$ is the fluid velocity and $P^{\mu\nu}=g^{\mu\nu}+u^\mu
u^\nu$ is the projection to the directions perpendicular to $u^\mu$.\footnote{We
assume the fluid is conformal so the bulk viscosity is zero.} 
On general grounds, there exists a
Hall viscosity contribution to the stress tensor
\begin{equation}\label{hall}
  \tau^{\mu\nu}_H= 
  - \frac12\eta_A (\epsilon^{\mu\alpha\beta} u_\alpha {V_\beta}^\nu 
  + \epsilon^{\nu\alpha\beta} u_\alpha {V_\beta}^\mu),
\end{equation}
which by construction, is only allowed in (2+1)-dimensions. It is
worth mentioning that the Hall viscosity term does not contribute to
entropy production and hence is dimensionless. In the comoving frame,
where the fluid velocity at a given point is $u^\mu=(1, \vec{0})$, the
Hall viscosity contribution to the stress tensor $\tau^{ij}$ has the
same form (up to a sign) as the one discussed in~\cite{Avron}
\begin{equation}
  \tau^{ij}_H = \frac12\eta^{ijkl}_A V_{kl},\qquad
  \eta^{ijkl}_A=-\frac12\eta_A
  (\delta^{ik}\epsilon^{jl}+\delta^{jk}\epsilon^{il}
  + \delta^{il}\epsilon^{jk}+\delta^{jl}\epsilon^{ik}). 
\end{equation}

In this paper we write down a holographic theory which realizes the
phenomenon of Hall viscosity. Our bulk theory is Anti de Sitter
gravity coupled to a gravitational Chern-Simons term. It is similar to
the model considered in~\cite{Jackiw}.  We find that the
corresponding boundary theory exhibits ``Hall viscosity.''  Moreover,
we discover that the Hall viscosity is completely determined by the
near horizon region of the black-brane.

The paper is organized as follows. In section two, we present our
holographic setup in detail. In section three, the Hall viscosity
contribution to the stress tensor is computed using fluid/gravity
correspondence. Section four is where a Kubo's formula is proposed and
is further used to calculate the Hall viscosity coefficient. We
conclude the presentation by the outlook section. Some details of the
holographic renormalization of our model appears in the appendix.
\section{The setup}
\label{sec:setup}
It is interesting to construct gravity backgrounds dual to isotropic
hydrodynamic flows with a non-vanishing Hall viscosity term. This is
only possible in (2+1)-dimensions and in the presence of a broken
time-reversal or parity.  Here we propose a gravity dual which
realizes this goal. We do not insist on deriving the gravity dual from
string theory in a top-down approach. Rather we take a more
phenomenological attitude. We introduce a parity-breaking interaction
by turning on a gravitational $\theta$-term in the bulk action. Parity
breaking alternative gravity theories have been studied in the past
\cite{Jackiw} with phenomenology in mind. We consider a generalization
of this class of models which includes a negative cosmological
constant.

Let us begin by describing the setup. Throughout the main text, we use the
uppercase Latin letters for the four spacetime coordinates. Lower case
letters $i, j$ are reserved for the spatial boundary field theory
directions. Greek letters refer to the boundary coordinates both
temporal and spatial. The convention we follow for the
$\epsilon$-tensor is $\epsilon^{vxy}=1$, where
$v$ is null and $x$ and $y$ are space-like. .

Our bulk theory lives in four spacetime dimensions. The Lagrangian
density is
\be\label{action}
\mathcal{L}=R-2\Lambda -\frac12(\partial\theta)^2-V(\theta)
-\frac\lambda4\theta ~^*\!R R,
\ee
where 
\begin{equation}
\begin{split}
  ^*\!RR&= \, ^*\!R^{M~\,PQ}_{~\,\,N}\,R^N_{~\,MPQ}, \\
  ^*\!R^{M~\,PQ}_{~\,\,N}&=\frac12\epsilon^{PQAB}R^M_{~\,\,N\!AB},
\end{split}
\end{equation}
and $\epsilon^{ABCD}$ is the four dimensional Levi-Civita tensor and
$\lambda$ is a coupling constant. We set
$\Lambda=-3$ from now on. Note that in order for the gravitational $\theta$-term to have a
nontrivial effect on the field equations, the field $\theta$ must be
spacetime-dependent. 

Varying the action (\ref{action}) with respect to $\theta$ and the
metric leads to the following field equations
\begin{equation}\label{eom}
\begin{split}
  G_{M\!N}+\Lambda g_{M\!N}-\lambda C_{M\!N}&= T_{M\!N}(\theta),\\
  \nabla^2\theta&= \frac{dV}{d\theta}+\frac\lambda4\, ^*\!R R,
\end{split}
\end{equation}
where 
\begin{equation}\label{cotton}
\begin{split}
  C^{MN}&=m_A\epsilon^{ABP(M}\nabla_{\!P}R^{N)}_{~\,B}
      +m_{AB}\,^*\!R^{B(M\!N)A}\,,\\
  T_{M\!N}&=\frac12\partial_M\theta\partial_N\theta
    -\frac14g_{M\!N}(\partial\theta)^2-\frac12g_{M\!N}V(\theta),
\end{split}
\end{equation}
and $m_{M}=\nabla_{\!M}\theta,
m_{M\!N}=\nabla_{\!M}\nabla_{\!N}\theta=\nabla_{(\!M}\nabla_{\!N)}\theta$. In the above the
symmetric-traceless $C$-tensor is the analog of the Cotton tensor in
three-dimensions. Similar equations of motion (\ref{eom}) were also
derived in for example~\cite{Jackiw, Grumiller}. Details of variation
of the action and holographic renormalization are gathered and briefly
discussed in the appendix~\ref{app}.

To write down a black-brane background solution of the equations of motion 
(\ref{eom}), we take the following ansatz
\begin{equation}\label{ansatz}
\begin{split}
ds^2=g^{(b)}_{M\!N}dx^{M}dx^{N}&=
  2H(r)dvdr-r^2f(r)dv^2+r^2dx\cdot dx,\\
\theta&= \theta^{(b)}(r).
\end{split}
\end{equation}
We note that \cite{Grumiller} for any ansatz of the above form, the
Pontryagin form $^{*}RR$ is identically {\it zero}. In addition to this,
for the ansatz (\ref{ansatz}), the $C$-tensor vanishes
identically. These two observations imply that any solution of the
form (\ref{ansatz}) to the following system
\begin{equation}\label{GR}
\begin{split}
  G_{M\!N}+\Lambda g_{M\!N}&=T_{M\!N}(\theta),\\
\nabla^{2}\theta&=\frac{dV}{d\theta},
\end{split}
\end{equation}
will give rise to a solution of our system (\ref{eom}).\footnote{We choose boundary conditions for $\theta$ such that $\theta$ is a relevant deformation. We never source $\theta$} We also  We write a
general formula for the Hall viscosity in terms of a general
background solution (\ref{ansatz}), so one is not required to be more
specific about the background solution. In passing, for future use
we record the Hawking temperature and the entropy density of the
black-brane (\ref{ansatz}) 
\be
T=\frac{r_{H}^2f'(r_{H})}{4\pi H(r_{H})}, \quad s=\frac{r_{H}^2}{4G_{4}}.
\ee 
\section{Fluid dynamics/gravity correspondence}
\label{sec:fluid-gravity}
In this section, we perform the fluid/gravity procedure as appears in
\cite{Shiraz} (see also \cite{Mark}). Before proceeding to the computation, let us outline briefly the algorithm.\footnote{We only need first order hydrodynamics to compute the Hall viscosity.} The idea is to systematically find the gravity
background describing the boundary hydrodynamics in a derivative
expansion. The background geometry
\ba
ds^2&=&-2H(r, b)u_{\mu}dx^{\mu}dr-r^2f(r, b)u_{\mu}u_{\nu}dx^{\mu}dx^{\nu}
  +r^2(\eta_{\mu\nu}+u_{\mu}u_{\nu})dx^{\mu}dx^{\nu},\\\nonumber
\theta&=&\theta(r,b),
\ea 
describes the boundary hydrodynamics in (2+1)-dimensions at thermal
equilibrium, where $b$ is the black-brane's Hawking temperature and
$u^{\mu}=(1{-}\vec{\beta}^2)^{-1/2}(1, \vec{\beta})$. If one promotes
$u^{\mu}$ and $b$ to slowly varying functions of the boundary coordinates,
the resulting inhomogeneous background (expanded up to first
derivative) will cease to be a solution
\begin{equation}\label{corr}
\begin{split}
ds^2_{(1)}&= 2H(r)dvdr-r^2f(r)dv^2+r^2dx\cdot dx 
  +\epsilon\bigl[-r^2\delta b~\partial_{b}f dv^2
     +2\delta b~\partial_{b}H~dvdr \\
  &\qquad -2r^2(1-f(r))x^{\mu}\partial_{\mu}\beta^{(0)}_{i}dvdx^{i}
  -2H(r) x^{\mu}\partial_{\mu}\beta^{(0)}_{i}drdx^i\bigr], \\
\theta^{(1)} &=\theta^{(b)}(r)
   +\epsilon [u^{\mu}\frac{\partial{\theta}}{\partial u^{\mu}}+\delta b
~\partial_{b}\theta],
\end{split}
\end{equation}
where derivatives are evaluated on the background. In the above
$\epsilon$ counts the number of derivatives along the boundary. The
procedure is to correct the resulting geometry by adding an extra piece of order $\epsilon$. We parametrize the correction as
\begin{equation}
\begin{split}
ds^2_{corr}&=
\epsilon\Bigl[\frac{k(r)}{r^2}dv^2+2h(r)dvdr-r^2h(r)dx\cdot dx
  +\frac{2}{r}a^{i}(r)dvdx^{i}+r^2\alpha_{ij}(r)dx^{i}dx^{j}\Bigr],\\
\theta_{corr}&=\epsilon\Theta(r),
\end{split}
\end{equation}
where $\alpha_{ij}$ is taken to be symmetric and traceless. The
trace-reversed form of the Einstein equation is more appropriate for
the fluid/gravity correspondence
\begin{equation}\label{EoMS}
\begin{split}
 E_{M\!N}&= R_{M\!N}+3g_{M\!N}-\lambda C_{M\!N}-d_{M\!N}=0,\\
 \nabla^2\theta&= \frac{dV}{d\theta}+\frac{1}{4} {}^{*}\!R R,
\end{split}
\end{equation}
where 
\be
d_{M\!N}=\frac{1}{2}(\partial_{M}\theta\partial_{N}\theta+g_{M\!N}V(\theta)).
\ee
Before proceeding, the following simplifying observation will prove helpful. It turns out that the general structure of the perturbation theory is as follows
\be\label{gf}
  \mathfrak{F}_{M \! N}[\epsilon h, g^{(b)}_{M\!N} ]
  =\epsilon \lambda C^{(1)}_{M\!N}(\theta^{(b)}\!,\, g^{(b)}_{M\!N})
  +\epsilon d^{(1)}_{M\!N},
\ee
where $\mathfrak{F}_{M\!N}$ is a linear differential operator
(containing only radial derivatives) acting on $h$, where $h$
collectively refers to first order gravity fluctuations. The
superscript ``(1)'' denotes first order (in $\epsilon$)
quantities. Also note that the $C$-tensor vanishes on the background
as previously mentioned. Evidently, the Hall viscosity term can only 
originate from $C^{(1)}_{M\!N}$. Here we are
only interested in computing the coefficient of Hall viscosity. Corrections (proportional to $\epsilon$) to $\theta^{(b)}$ will generate higher order terms in
$\epsilon$ on the right-hand side of (\ref{gf}). Similarly, corrections (proportional to $\epsilon$) to $g^{(b)}_{M\!N}$ on the left hand side produce terms which are irrelevant to the Hall viscosity computation. Therefore, as long as first order hydrodynamics is concerned, we can only use the background solution for $\theta$ and $g_{M\!N}$.

As in~\cite{Shiraz,Mark}, various components of the equations of
motion correspond to the constitutive relations and/or the hydrodynamic
equations of motion. The Hall viscosity coefficient can be computed
from $T_{xy}$ (and/or $T_{xx}-T_{yy}$) component(s) of the stress
tensor. For this, studying the tensor sector will suffice. Along the way, we observe that
\be
C^{(1)}_{xy}=\frac1{4H}\frac d{dr}
  \Bigl(\frac{r^4f' \theta'}{H^2}\Bigr)
  (\partial_{x}\beta_{x}-\partial_{y}\beta_{y}), \quad 
C^{(1)}_{xx}-C^{(1)}_{yy}=\frac1{2H}\frac d {dr}\Bigl(
   \frac{r^4f' \theta'}{H^2}\Bigr)
  (-\partial_x\beta_y-\partial_y\beta_x).
\ee
From $E^{(1)}_{xy}=0$
\begin{equation}\label{c}
\begin{split}
\frac1H\frac d {dr}\Bigl[-\frac12\frac{r^4f(r)}{H(r)}
  \frac d{dr}\alpha_{xy}(r)\Bigr]
  -2\frac rH\sigma_{xy}&\\
+[\frac{r^3H'f}{H^3}-r^3\frac{f'}{H^2}-3r^2\frac{f}{H^2}+3r^2
-\frac{r^2}{2}V(\theta)]\alpha_{xy}(r) &=
  \frac\lambda{4H}\frac d {dr}\Bigl(\frac{r^4f' \theta'}{H^2}\Bigr)
  (\partial_x\beta_x-\partial_y\beta_y).
\end{split}
\end{equation}
Background equations of motion imply
\be\label{bg}
E^{(b)}_{xx}=0=\frac{r^3H'f}{H^3}-r^3\frac{f'}{H^2}
-3r^2\frac{f}{H^2}+3r^2-\frac{r^2}{2}V(\theta).
\ee
Putting Eqs.~(\ref{c}) and (\ref{bg}) together, for a general
background solution (\ref{ansatz}), one obtains
\be\label{tensor1}
\frac1H\frac d {dr}\Bigl[-\frac12\frac{r^4f(r)}{H(r)}
  \frac d{dr}\alpha_{xy}(r)\Bigr]
=2\frac rH\sigma_{xy}+\frac\lambda{4H}\frac d {dr}
  \Bigl(\frac{r^4f' \theta'}{H^2}\Bigr)
  (\partial_x\beta_x-\partial_y\beta_y),
\ee
where 
\be
\sigma_{ij}=\frac12(\partial_i\beta_j+\partial_j\beta_i)
  -\frac12\delta_{ij}\partial_k\beta_k.
\ee
Also from $E^{(1)}_{xx}-E^{(1)}_{yy}=0$
\ba\label{tensor2}
\frac1H\frac d{dr}\Bigl[-\frac12\frac{r^4f(r)}{H(r)}\frac d{dr}
  \alpha_{xx}(r)\Bigr]
=2\frac rH\sigma_{xx}+\frac\lambda{4H}\frac d{dr}
\Bigr(\frac{r^4f' \theta'}{H^2}\Bigr)(-\partial_x\beta_y-\partial_y\beta_x).
\ea
Having solved for $\alpha_{xx}$ from the above, using the traceless
condition we have
\be\label{rel}
\alpha_{yy}(r)=-\alpha_{xx}(r). 
\ee
\subsection{Tensor perturbations}
Let us concentrate on equations (\ref{tensor1}) and (\ref{tensor2}).
One can integrate (\ref{tensor1}) and solve for $\alpha_{xy}$
\be\label{tensorxy_sol}
\alpha_{xy}(r)=\int^{\infty}_{r} \frac{2H(s)ds}{s^4f(s)}\int^s_{r_{H}}\! dw\, 
  \mathfrak{S}_{xy}(w),
\ee
where 
\be
\mathfrak{S}_{xy}(r)=2r\sigma_{xy}+\frac\lambda4\frac d{dr}
\Bigl(\frac{r^4f' \theta'}{H^2}\Bigr)(\partial_x\beta_x-\partial_y\beta_y),
\ee
where the source free solution is discarded since it is not
normalizable. The following formula is useful when we compute the
stress tensor
\begin{equation}
\begin{split}
r^n\alpha_{xy}(r)&\rightarrow -\frac{r^{n+1}}n\frac{d\alpha_{xy}(r)}{dr},\\
r&\rightarrow\infty.
\end{split}
\end{equation}
In order to write a general formula we further assume 
\be
f(r)=1-\mathcal{O}({r_H}^3/r^3),\quad H(r)=1-\mathcal{O}({r_{H}}^3/r^3).
\ee
These assumptions are not necessarily the minimal requirements
enabling one to write a general formula for the Hall viscosity. With these assumptions,
one can write the contribution of the Hall term to the stress tensor
\be\label{xy}
T^{Hall}_{xy}=-\frac\lambda{8\pi G_{4}}
  \frac{r^4f'(r)\theta'(r)}{4H^2(r)}\bigg|_{r=r_H}
  (\partial_x\beta_x-\d_y\beta_y).
\ee
Similarly, integrating (\ref{tensor2}), one obtains 
\be
\alpha_{xx}(r)=\int^{\infty}_r \frac{2H(s)ds}{s^4f(s)}
  \int^s_{r_H}\! dw\, \mathfrak{S}_{xx}(w),
\ee
where
\be
\mathfrak{S}_{xx}(r)=2r\sigma_{xx}
  +\frac\lambda4\frac d {dr}\Bigl(\frac{r^4f' \theta'}{H^2}\Bigr)
  (-\partial_x\beta_y-\partial_y\beta_x).
\ee 
The gravity stress tensor will give the Hall contribution 
\be\label{xx_yy}
T^{Hall}_{xx}-T^{Hall}_{yy}=\frac\lambda{8\pi G_{4}}
  \frac{r^4 f'(r)\theta'(r)}{2H^2(r)}\bigg|_{r=r_H}
  (\partial_x\beta_y+\partial_y\beta_x).
\ee 
Comparing the definition of the Hall viscosity (\ref{hall}) and our
results (\ref{xx_yy}) and (\ref{xy}), we obtain
\be
  \eta_A=-\frac\lambda{8\pi G_{4}}
  \frac{r^4f'(r)\theta'(r)}{4H^2(r)}\bigg|_{r=r_H}.
\ee
This is a general membrane-paradigm-type formula~\cite{MP} for the
Hall viscosity coefficient, in the sense that the Hall viscosity is entirely determined by the near
horizon region of the black-brane geometry.
 \section{A Kubo's formula for $\eta_A$}
\label{sec:Kubo}
Hall viscosity can also be computed at zero spatial momentum. Generally speaking, a
transport coefficient could be viewed as parametrizing the response of
a fluid to hydrodynamic perturbations. One way to induce these
disturbances is by perturbing the non-dynamical boundary metric
$g_{\mu\nu}=\eta_{\mu\nu}+h_{\mu\nu}+\mathcal{O}(h^2) $. We work in the local
rest frame of the fluid $u^{\mu}=(1, \vec{0})$ and at zero
spatial momentum. The minimal set of background metric perturbations which
needs to be switched on to compute the Hall viscosity is $h=
\{h_{xy}(t), h_{xx}(t),h_{yy}(t)\}$. One gets
\be\label{kubo}
T^{xy}=-P h_{xy}-\eta\frac{\partial h_{xy}}{\partial t}
 +\frac{\eta_A}2\frac{\partial}{\partial t}(h_{xx}-h_{yy}),
\ee 
where $P$ is the pressure. A similar formula for the shear viscosity
has been written down before; see for example~\cite{Son}. Here
we have the extra term proportional to the Hall viscosity.
\subsection{Linearized gravity perturbation}
In this section we study linearized gravitational perturbations of the
background (\ref{ansatz}). We write down equations of motion
(\ref{eom}) for the following linearized perturbations
\be 
g_{M\!N}=g^{(b)}_{M\!N}+ g^{(1)}_{M\!N}, \quad g^{(1)}_{xy}=
r^2h_{xy}(r)e^{-i\omega v}, \quad g^{(1)}_{xx}=r^2h_{xx}(r)e^{-i\omega
v}, \quad g^{(1)}_{yy}=r^2h_{yy}(r)e^{-i\omega v},
\ee
where $h_{xx}(r)=-h_{yy}(r)$. Note that in this coordinate system, the
horizon is a non-singular surface. Boundary condition at the horizon
is incoming, since we are interested in the response. At infinity we
impose $h_{yy}(r)\rightarrow H^{0}_{yy}, h_{xy}(r)\rightarrow
H^{0}_{xy}$. The equations of motion are solved in an expansion in small frequencies. It turns out that one can solve the equations for a general background. We only care about the first order correction in $\omega$. One obtains
\begin{equation}
\begin{split}
h_{xx}(r)&=H^0_{xx}+\frac\omega2\int^r_\infty
\frac{-i\lambda H^0_{xy} s^4 f'(s)\theta'(s)+2iH^0_{xx}s^2H^2(s)
  +2c_1 H^2(s) }{s^4f(s)H(s)}\,ds ,\\
h_{xy}(r)&=H^0_{xy}+\frac\omega2\int^r_\infty\frac{i\lambda H^0_{xx} 
  s^4 f'(s)\theta'(s)+2iH^0_{xy}s^2H^2(s)+2c_2 H^2(s) }{s^4f(s)H(s)}\,ds,
\end{split}
\end{equation}
where the integration constants $c_1$ and $c_2$, are determined through
demanding regularity for $h_{xx}$ and $h_{xy}$ at the horizon. This
leads to
\be
c_1=-iH^0_{xx}r_H^2+i\lambda H^0_{xy}~ \frac{s^4f'(s)\theta'(s)}{2H^2(s)}
  \bigg|_{s=r_{H}}, \quad 
c_2=-iH^0_{xy}r_H^2-i\lambda H^0_{xx}~\frac{s^4f'(s)\theta'(s)}{2H^2(s)}
  \bigg|_{s=r_{H}}.
\ee
As shown in the appendix, the Chern-Simons term does not contribute to the stress tensor. Let us
focus on the $xy$-component of the stress tensor. The relevant (to the
Hall viscosity computation) part of $h_{xy}(r)$ is
\be
  h_{xy}(r)= H^0_{xy}+\frac{i\lambda\omega}{3r^3}W H^{0}_{xx}+\cdots,
\ee
where 
\be
W=\frac{r^4 f'(r)\theta'(r)}{2H^2(r)}\bigg|_{r=r_{H}}.
\ee 
The gravity stress tensor gives 
\be
8\pi G_{4}T_{xy}=\frac{i\lambda\omega}{2}WH^{0}_{xx},
\ee
Comparing this with the Kubo's formula (\ref{kubo}), we can read off the
Hall viscosity coefficient as
\be\label{Hallvisco}
\eta_A= - \frac\lambda{8 \pi G_{4}}\frac{r^4 f'(r)\theta'(r)}{4H^2(r)}
  \bigg|_{r=r_{H}}.
\ee 
This is exactly equal to what we computed using
fluid/gravity duality in the previous section. 

\section{Outlook}
\label{sec:outlook}
In this paper, we have constructed a holographic model
exhibiting Hall viscosity.  Although a non-zero Hall
viscosity coefficient was not unexpected given the choice of interactions in our model, it is gratifying to find that the rules of gauge/gravity duality
indeed lead to a Hall viscosity.  It is interesting to note that the
value of the Hall viscosity, in our model, depends only on the
behavior of the scalar field $\theta$ at the horizon, which indicates
that there exists a membrane paradigm principle that fixes the value
of this kinetic coefficient at the black-brane horizon. It would be
interesting to find out if that is true and how the Hall viscosity is
communicated to the boundary (see, e.g., \cite{Nickel:2010pr}).  It
is also interesting to explore the connection between the holographic
approach developed here with the purely Lagrangian approach to Hall
viscosity of~\cite{Nicolis:2011ey}.
\section*{Acknowledgement}
\label{sec:ack}
We thank K.~Jensen, M.~Kaminski, R.~Myers, and A.~Yarom for discussions. 
O.S. would like to thank INT/University of Washington for
hospitality and also Petr Ho\u{r}ava, Alberto Nicolis and Kevin Schaeffer for discussions.
The work of
D.T.S. is supported, in part, by DOE grant No.\
DE-FG02-00ER41132.  O.S. is supported by the Berkeley Center for Theoretical Physics,  department of physics at UC Berkeley and in part by DOE, under contract 
DE-AC02-05CH11231. 
\begin{appendix}
\section{Holographic renormalization}
\label{app:hol-ren}
\label{app}
In this appendix lower case Latin indices $a, b, c, \cdots$ run over all four spacetime coordinates. The Latin indices $i, j, k, \cdots$ refer to the spatial coordinates. We set $16\pi G_{4}=1$. To compute the Hall viscosity contribution to the hydrodynamic flow of the boundary theory, one has to write down the stress-tensor associated with the theory (\ref{action}). It is also crucial to make sure that the action principle is well-defined. This is to say, one should only be required to keep fields (and not their normal derivatives) fixed at the boundary. The general procedure is to add the analog of a Gibbons-Hawking term to the bulk action (\ref{action}). For the Chern-Simons modified gravity (\ref{action}), this term has been computed in \cite{Grumiller}.
Here we show that for the particular case of interest in this paper, i.e., when $\theta$ is a relevant deformation (vanishes asymptotically), on any solution to the equations of motion coming from the action (\ref{action}), the stress-tensor is just that of asymptotically AdS$_4$ spaces. The only counter-terms needed are those of asymptotically AdS$_4$ spacetimes. Variation of the Chern-Simons term is straightforward
\ba\label{vari}
\delta S_{CS}&=&-\frac{\lambda}{4}\delta\int d^4x \sqrt{-g} \theta ^{*}RR,\\\nonumber
&=&-\lambda\int d^4x\sqrt{-g}\nabla_{c}(\theta ^{*}R_{~a}^{b~cd}\delta\Gamma^{a}_{~bd} )+\lambda \int d^4x \sqrt{-g}\nabla_{b}[\delta g_{ed}\nabla_{c}(\theta~ ^{*}R^{becd})]-\lambda S_{1},\\\nonumber
S_{1}&=&\int d^4x \sqrt{-g}\delta  g_{ed}\nabla_{b}\nabla_{c}(\theta~ ^{*}R^{becd}).
\ea
On the second line of the above equation, there are two boundary terms which we come back to later. For the moment let us focus on $S_{1}$. 
Using the second Bianchi identity $R^{be}_{~~[fg;c]}=0$, its contracted form $R^{be}_{~~fg;b}=R^{e}_{~g;f}-R^{e}_{~f;g}$ and the fact that $\delta g_{ed}$ is symmetric, $S_1$ can be rewritten as  
\be
S_{1}=-\int d^4x \sqrt{-g} \delta g_{ed}[\nabla_{(b}\nabla_{c)}\theta~^{*}R^{c(ed)b}+\nabla_{c}\theta \epsilon^{cgf(d}\nabla_{f}R^{e)}_{~g} ]=-\int d^4x \sqrt{-g} \delta g_{ed} C^{ed},
\ee
where the $C$-tensor (\ref{cotton}) definition was utilized. Now let us collect all the boundary terms on the second line of (\ref{vari})
\be\label{deltaS}
\delta S_{b}=-\lambda\int d^4x\sqrt{-g}\nabla_{c}(\theta ^{*}R_{~a}^{b~cd}\delta\Gamma^{a}_{~bd} )+\lambda\int d^4x \sqrt{-g}\nabla_{b}[\delta g_{ed}\nabla_{c}(\theta~ ^{*}R^{becd})].
\ee
We show that the $\theta$ dependent boundary terms above will vanish when $\theta$ is a relevant perturbation. We were able to demonstrate this in the Gaussian normal coordinates
\be\label{Ng}
ds^2=dr^2+g_{ij}dx^{i}dx^{j},
\ee
using standard identities (see \cite{Kraus} for example). Let us focus on the first boundary term in (\ref{deltaS}) 
 \be
S_{2}=-\lambda\int d^4x\sqrt{-g}\nabla_{c}(\theta ^{*}R_{~a}^{b~cd}\delta\Gamma^{a}_{~bd} )=\lambda\int d^3x \theta( 2\varepsilon^{r j k l}\nabla_{k}K^{i}_{~l}\delta K_{ij}+\sqrt{-h}~^{*}R_{j}^{~k r i}
\delta\Gamma^{j}_{ki}),
\ee
where Codazzi equation in the coordinates (\ref{Ng}) 
\be
^{*}R_{a}^{~b r d}\delta\Gamma^{a}_{bd}=^{*}R_{j}^{~k r i}\delta\Gamma^{j}_{ki}+2~^{*}R_{r}^{~j r i}\delta\Gamma^{r}_{ji},
\ee
as well as the following two identities (true in the normal coordinates), $R^{i r}_{~kl}=^{(3)}\nabla_{k}K^{i}_{~l}-^{(3)}\nabla_{l}K^{i}_{~k}$ and $K_{ij}=\frac{1}{2}\frac{\partial g_{ij}}{\partial r}$ were used. Note that the Fefferman-Graham expansion of the metric and extrinsic curvature are given by $g_{ij}=e^{2r}g^{(0)}_{ij}+g^{(2)}_{ij}+\cdots$,
$K_{ij}=e^{2r}g^{(0)}_{ij}+0+\cdots$ and
$K^{i}_{~j}=\delta^{i}_{~j}-e^{-2r}g^{il}_{(2)}g^{(0)}_{lj}+\cdots$. Utilizing these, one concludes that $S_2$ vanishes at least as fast as   
\ba
S_{2}&\sim &\int_{\partial} d^3x ~\theta \rightarrow 0.
\ea
The argument showing that the second $\theta$-dependent boundary term also vanishes proceeds similarly. The last term in (\ref{deltaS}) can be further simplified (using the second Bianchi identity)
\ba
S_{3}&=& \frac{\lambda}{2}\int d^3x  \varepsilon^{r d f g}\partial_{r}\theta R^{r e}_{~~fg}\delta g_{ed},\\\nonumber
&=&-\lambda\int d^3 x \varepsilon^{rdfg}\partial_{r}\theta \nabla_{f} K^{e}_{~g}\delta g_{ed}
\rightarrow 0.
\ea
In sum, the only terms relevant to the variation of the action (including the conventional Gibbons-Hawking term) are
\ba
\delta S=-\int d^4x \sqrt{-g} \delta g_{ed}(G^{ed}+\Lambda g^{ed}-\lambda C^{ed}) -\int d^3x \sqrt{h}(K^{ed}-h^{ed}K)\delta g_{ed}.
\ea
The only counter-term will be a boundary cosmological constant renormalization since our boundary is flat. The stress tensor is then 

\be
\delta S=\frac{1}{2}\int d^3x \sqrt{g^{(0)}_{ij}} T^{ij}\delta g^{(0)}_{ij}=-\int  d^3x \sqrt{h}(K^{ij}-g^{ij}K-2g^{ij})\delta g_{ij}.
\ee

The background can be written in an ADM form
\ba
ds^2&=&N^2dr^2+g_{\alpha\beta}(dx^{\alpha}+N^{\alpha}dr)(dx^{\beta}+N^{\beta}dr),\\\nonumber
&=& (N^2+g_{\alpha\beta}N^{\alpha}N^{\beta})dr^2+2N_{\alpha}dx^{\alpha}dr+g_{\alpha\beta}dx^{\alpha}dx^{\beta},
\ea
where $N_{\alpha}=g_{\alpha\beta}N^{\beta}$. In our gauge $g_{rr}=0$ so $N^{2}+N_{\alpha}N^{\alpha}=0$. The extrinsic curvature is calculated using the standard formula
\be
K_{\alpha\beta}=\frac{1}{2N}(N_{\alpha;\beta}+N_{\beta;\alpha}-\frac{\partial g_{\alpha\beta}}{\partial r}).
\ee

\end{appendix}

\end{document}